\documentclass[a4paper,11pt]{article}
\usepackage{jheppub}
\usepackage{eurosym}
\usepackage{amssymb}
\usepackage{amsfonts,cite}
\usepackage{amsmath}
\usepackage{graphicx}
\usepackage{multirow}
\usepackage{xcolor}
\usepackage{epsfig}
\usepackage{fancyhdr}

\newbox\mybox

\newcommand\fverb{\setbox\mybox=\hbox\bgroup\verb}
\newcommand\fverbdo{\egroup\medskip\noindent\fbox{\unhbox\mybox}\ }
\newcommand\fverbit{\egroup\item[\fbox{\unhbox\mybox}]}

\abstract{We investigate a three-dimensional ghostly Hamiltonian realisation of the fully degenerate resonant sixth-order Pais-Uhlenbeck oscillator. On the classical level, the phase-space flow is non-diagonalisable and decomposes into two complex-conjugate Jordan chains of length three, explaining the appearance of oscillatory solutions with secular terms.
	
	Upon quantisation, we construct intertwining operators whose quadratic combinations generate a hidden spectrum-generating $\mathfrak{u}(2,1)$-algebra. The associated descendant spaces are finite-dimensional invariant subspaces carrying non-trivial Jordan structure. Although these spaces admit a natural decomposition into irreducible modules of a distinguished $\mathfrak{sl}_2$-subalgebra, this decomposition does not in general coincide with the Jordan decomposition of the Hamiltonian.
	
	We further derive a tri-Hamiltonian formulation from Lie point symmetries of the classical flow and show that the corresponding Hamiltonians are naturally encoded by the same hidden algebra. Nevertheless, unlike in the non-resonant case, no positive-definite linear combination of them generates the same dynamics. Finally, we analyse the common centraliser of the tri-Hamiltonian family in $U(\mathfrak u(2,1))$, showing that the natural higher-order candidate $Q$ is reducible and yields no independent classical or quantum integral.
	
	The model thus provides a resonant higher-derivative system in which hidden $\mathfrak{u}(2,1)$ symmetry, classical and quantum Jordan structures, and multi-Hamiltonian geometry coexist.}  
	
	\author[a]{Andreas Fring,}
\author[b]{Ian Marquette,}

\title{Hidden $\mathfrak{u}(2,1)$ symmetry and Jordan chains in a resonant ghostly three-dimensional model}

\affiliation[a]{Department of Mathematics, City St George's, University of London,  Northampton Square, \\ London EC1V 0HB, UK}
\affiliation[b]{Department of Mathematical and Physical Sciences, La Trobe University, Bendigo, VIC 3552, Australia}

\emailAdd{a.fring@city.ac.uk}	
\emailAdd{i.marquette@latrobe.edu.au}

\begin{document}
	\maketitle
	
	\pagestyle{fancy}
	\fancyhead{} % clear all header fields
	\fancyhead[LE,RO]{\small\itshape  Hidden $\mathfrak{u}(2,1)$ symmetry and Jordan chains in a resonant ghostly 3D model} 
	
	\renewcommand{\headrulewidth}{0.4pt}

\section{Introduction}

Higher time-derivative theories (HTDTs) occur in effective field theory \cite{buchbinder2017effective}, modified
gravity \cite{stelle77ren,starobinsky1980new,sotiriou2010f} and regularisation schemes \cite{stelle77ren,grav3,tomboulis2015renormalization,modesto2012super,modesto2016superrenormalizable,barnaby2008dynamics}, but they are typically accompanied by
Ostrogradsky instabilities \cite{ghostconst,motohashi1,motohashi4} and ghost degrees of freedom \cite{Woodard1,weldon03quant,smilga2017class}. The  Pais-Uhlenbeck (PU) \cite{pais1950field} 
oscillator is the simplest model in which these issues can be studied
explicitly. It therefore provides a useful testing ground for understanding
whether hidden algebraic structures, alternative Hamiltonian formulations and
non-standard quantisation schemes can control or reorganise the ghostly sectors
of higher time-derivative dynamics.

In \cite{fring2026spectrum} we analysed the resonant PU
oscillator in its ghostly two-dimensional Hamiltonian formulation. At the
equal-frequency point the usual Fock-space construction collapses, the
Hamiltonian becomes non-diagonalisable, and the quantum states are naturally
organised into finite Jordan chains rather than ordinary eigenspaces. The
central result of that analysis was the emergence of a hidden
spectrum-generating $\mathfrak{su}(2)$ algebra, generated by quadratic
combinations of intertwining operators. For the real models considered this algebra exists only at resonance
and provides an algebraic mechanism for organising the generalised
eigenvectors of the fourth-order degenerate PU model. For systems with complex interactions hidden algebras have been identified in \cite{marquette22lad1,marquette22lad2}.

Here we continue this programme by moving from the fourth-order to
the sixth-order fully degenerate PU system. We consider a ghostly
three-dimensional Hamiltonian with one negative kinetic-energy direction and
a distinguished set of quadratic couplings. For the parameter choice studied
below, the three-dimensional equations of motion reduce to the threefold
degenerate sixth-order PU equation
\begin{equation}
\left(\frac{d^2}{dt^2}+\omega^2\right)^3 q(t)=0 . \label{6PU}
\end{equation}
Thus the resonant behaviour found in the fourth-order model is enhanced:
besides the usual oscillatory factors, the classical solutions contain
polynomial factors $t$ and $t^2$. Equivalently, the first-order classical
flow is no longer described by two $2\times2$ Jordan blocks, but by two
$3\times3$ Jordan blocks associated with the eigenvalues
$\pm i\omega$.

This higher degeneracy raises the natural question of whether the hidden
$\mathfrak{su}(2)$ structure of the fourth-order model persists, and if so
in what form. We show here that the answer is affirmative, but with a substantial
enlargement of the algebra. From three pairs of first-order intertwining
operators we construct quadratic generators which close into a
nine-dimensional Lie algebra isomorphic to $\mathfrak u(2,1)$. This algebra
contains a central element, a distinguished $\mathfrak{sl}_2$-subalgebra,
and a five-dimensional irreducible $\mathfrak{sl}_2$-module. In this sense,
the hidden algebra of the fourth-order resonant model is not an isolated
phenomenon, but the first member of a richer algebraic hierarchy associated
with higher-order resonant PU systems.

At the quantum level we construct a formal vacuum state annihilated by the
lowering intertwiners. As in the fourth-order case, this vacuum is not
normalisable in the standard $L^2$-sense, reflecting the ghostly nature of
the resonant sector. Nevertheless, it provides a useful algebraic starting
point. Acting with the raising operators generates finite-dimensional
descendant spaces
\begin{equation}
\mathcal V_N=
\operatorname{span}\{
a_+^\ell c_+^m b_+^n\psi_0:\ell+m+n=N\}.
\end{equation}
Each $\mathcal V_N$ is invariant under the Hamiltonian and forms a
generalised eigenspace with eigenvalue $2\lambda N$. The nilpotent part of
the Hamiltonian acts non-trivially on these spaces and produces finite Jordan
chains, now of greater length than in the fourth-order model.

The hidden $\mathfrak u(2,1)$ structure also reveals a natural
$\mathfrak{sl}_2$-module decomposition of the descendant spaces. Since
$\mathcal V_1$ is a three-dimensional irreducible $\mathfrak{sl}_2$-module,
the degree-$N$ descendant space is naturally identified with
$\operatorname{Sym}^N(D_2)$, and hence decomposes as
\begin{equation}
\mathcal V_N
\cong
\bigoplus_{k=0}^{\lfloor N/2\rfloor} D_{2N-4k}. \label{decompVN}
\end{equation}
This decomposition gives a compact representation-theoretic description of
the resonance sector. However, a new feature appears in comparison with the
simpler fourth-order case: the $\mathfrak{sl}_2$-module decomposition does
not in general coincide with the Jordan decomposition of the Hamiltonian.
The Hamiltonian is not an $\mathfrak{sl}_2$-intertwiner, and its nilpotent
part may mix different irreducible $\mathfrak{sl}_2$-summands. Thus the
hidden algebra organises the representation content of the theory, while the
actual Jordan normal form requires an additional analysis.

We also extend the Hamiltonian analysis of the resonant model. In the
two-dimensional fourth-order case, the resonant system admits more than one
Hamiltonian formulation generating the same classical flow, leading to a
quantisation ambiguity \cite{dam2006,FFT}. Here we show that the sixth-order ghostly model
possesses an analogous but richer structure: by using linear Lie point
symmetries of the classical flow, we construct a tri-Hamiltonian formulation \cite{tri1,tri2,tri3}.
Three distinct Hamiltonians, together with three constant Poisson tensors,
generate the same equations of motion. Moreover, the quantum versions of
these Hamiltonians are naturally expressed in terms of the same hidden
$\mathfrak u(2,1)$ generators, showing that the spectrum-generating algebra
and the multi-Hamiltonian structure are closely intertwined. 

In addition, we analyse the common centraliser of the tri-Hamiltonian family in $U(\mathfrak u(2,1))$. A natural higher-order candidate arises from the centraliser of the abelian subspace $\langle v_1,v_2\rangle$ and indeed commutes with all three Hamiltonians $H_1,H_2,H_3$. However, this charge is not independent, as it can be written as a quadratic polynomial in the Hamiltonians themselves. Its classical commutative symbol is similarly dependent on the classical tri-Hamiltonians and hence does not provide an additional integral of motion. Consequently, the apparent enhanced quantum symmetry is reducible, and no genuinely new independent quantum charge is obtained at this order.

Finally, we prove a no-go result for positive-definite Hamiltonians in the
fully resonant sixth-order model. In the corresponding non-resonant
three-dimensional system, suitable linear combinations of the tri-Hamiltonians
can generate the same flow while yielding a positive-definite Hamiltonian \cite{felski2026three}.
We show that this mechanism fails at the fully degenerate point. For every
linear combination of the three Hamiltonians that still reproduces the
original flow, the kinetic and potential quadratic forms fail Sylvester's
criterion for positive-definiteness. Thus the obstruction already observed in
the fourth-order resonant case persists, and becomes even more rigid, in the
sixth-order fully degenerate model.

Our manuscript is organised as follows: In section 2 we derive the classical
Jordan-chain structure of the ghostly three-dimensional model and show its
equivalence to the threefold degenerate sixth-order PU equation. In section 3 we
construct the hidden spectrum-generating $\mathfrak u(2,1)$ algebra from
quadratic combinations of intertwining operators. In section 4 we analyse the
quantum descendant spaces, their $\mathfrak{sl}_2$-module decomposition and
their Jordan-chain structure under the Hamiltonian. In section 5 we derive
the tri-Hamiltonian formulation from Lie point symmetries of the classical
flow and relate the alternative Hamiltonians to the hidden algebra. In Section 6 we analyse the common centraliser of the tri-Hamiltonian family in $U(\mathfrak u(2,1))$, showing that the natural higher-order candidate $Q$ is reducible and gives no independent classical or quantum integral. In Section 7 we study how the tri-Hamiltonian partners and $Q$ act on the $\mathfrak{sl}_2$-adapted module structure, clarifying how they preserve natural filtrations while mixing irreducible summands and leading to distinct Jordan decompositions. In section 8 we prove that no positive-definite Hamiltonian can be obtained from linear combinations of the tri-Hamiltonians while preserving the same resonant flow. Our conclusions are stated in section 9.

\section{Classical Jordan chains  of the ghostly model at PU resonance}
We consider the three-dimensional Hamiltonian 
\begin{equation}
	{\cal H}(x,y,z,p_x, p_y,p_z) = \frac{a_x}{2} p_x^2 +  \frac{a_y}{2} p_y^2 +  \frac{a_z}{2} p_z^2 +  \frac{b_x}{2} x^2 +  \frac{b_y}{2} y^2 +  \frac{b_z}{2} z^2  + g_1 x y  +g_2 x z + g_3 y z .   \label{genLxy}
\end{equation}
Systems of this type are of particular recent interest for parameter choices
that give rise to a ``ghostly'' model in which the kinetic term is of Lorentzian type and the potential has a saddle point nature \cite{deffayet22ghost,deffayet23global,diez2024foundations,deffayet2026unitary}. Here we consider the specific parameter choices
\begin{equation}
      a_x=-a_y=a_z =2,\,\, b_x= 2(g^2+\lambda^2), \,\, b_y= 2(g^2-\lambda^2), \,\,
      b_z= 2 \lambda^2, \,\, g_1=2 g^2,\,\, g_2=g_3= - 4 g \lambda.
\end{equation}
This choice is of particular interest as it corresponds to the fully degenerate resonant point of the sixth-order PU-model as we now show. Our goal is to expose the hidden symmetry responsible for the spectral organisation, in analogy with the construction in \cite{fring2026spectrum} for the fourth-order PU-model.

Introducing the phase-space vector $\vec{q}=(x,y,z,\dot{x},\dot{y},\dot{z})^\intercal$, we first note, by using Hamilton's equations, that the equations of motion resulting from (\ref{genLxy})
\begin{equation}
	\ddot{x}= 8 g \lambda  z -4 \lambda ^2 x-4 g^2 (x+y),  \quad
	\ddot{y}=  4 g^2 (x+y)-8 g \lambda  z- 4 \lambda ^2 y , \quad
	\ddot{z}= 8 g \lambda(x+y) - 4 \lambda^2  z  , \label{thresecord}
\end{equation}
 can be written in the form
\begin{equation}
	 \dot{ \vec{q} } = A \vec{q}, \qquad   A= 
	 \begin{pmatrix}
	 0 & 0 & 0 & 1 & 0 & 0 \\
	 0 & 0 & 0 & 0 & 1 & 0 \\
	 0 & 0 & 0 & 0 & 0 & 1 \\
	 -4 \left(g^2+\lambda ^2\right) &  -4 g^2 &  8 g \lambda  & 0 & 0 & 0 \\
	 4 g^2 & 4 \left( g^2 - \lambda ^2  \right) &  -8 g \lambda  & 0 & 0 & 0 \\
	 8 g \lambda  & 8 g \lambda  & -4 \lambda ^2 & 0 & 0 & 0 
	 \end{pmatrix} .
\end{equation}
Thus, the solutions are given by the matrix exponential $\vec{q}(t) =e^{ t A } \vec{q}(0) $. 
The characteristic polynomial for $A$ is computed to
\begin{equation}
\chi_A(\mu)=\det(A- \mu \mathbb{I}  )=(4\lambda^2+\mu^2)^3,
\end{equation}
so that the eigenvalues $\mu=\pm 2 i \lambda$ have multiplicity three. For generic
$g\neq 0$, the matrix $A$ admits only one eigenvector for each of these eigenvalues, and is therefore non-diagonalisable. Instead we find a Jordan chain structure. Defining the vectors
\begin{equation}
v_0 = \left(    \frac{i}{2 \lambda },-\frac{i}{2 \lambda },0,-1,1,0\right)^\intercal, \qquad
v_1 =  \left(  -\frac{1}{4 \lambda ^2},\frac{1}{4 \lambda ^2},-\frac{1}{4 g \lambda },0,0,-\frac{i}{2 g}\right)^\intercal,  
\end{equation}
\begin{equation}	
v_2 =   \left(    -\frac{i \left(g^2+\lambda ^2\right)}{8 g^2 \lambda ^3},\frac{i}{8 \lambda ^3},-\frac{i}{8 g \lambda ^2},\frac{1}{4 g^2},0,0\right)^\intercal ,
\end{equation}
we have
\begin{equation}
 \left( A - 2 i \lambda \mathbb{I}_6   \right)v_0 =0, \qquad
  \left( A - 2 i \lambda \mathbb{I}_6   \right)v_1 =v_0, \qquad
   \left( A - 2 i \lambda \mathbb{I}_6   \right)v_2 =v_1.
\end{equation}
Introducing the matrix $U=(v_0,v_1,v_2,v_0^*,v_1^*,v_2^*)$, the flow matrix $A$ is similar to the direct sum of two $3\times 3$ Jordan blocks
\begin{equation}
A=UJU^{-1}, \qquad J=J_+^{(3)}\oplus J_-^{(3)}, \qquad
J_\pm^{(3)}=\pm 2i\lambda\,I_3+N,  \quad N=
\begin{pmatrix}
	0&1&0\\
	0&0&1\\
	0&0&0
\end{pmatrix},
\end{equation}
with $N^3=0$. The nontrivial nilpotent part is responsible for the secular terms, as is seen directly from the matrix exponential. It takes the form $e^{ t A } = U e^{t J} U^{-1}$, where in each sector we have
\begin{equation}
e^{tJ_\pm}=e^{\pm 2i\lambda t}e^{tN}
=e^{\pm 2i\lambda t}\left(I+tN+\frac{t^2}{2}N^2\right).
\end{equation}
Thus, besides the oscillatory factors $e^{\pm 2i\lambda t}$, the nontrivial nilpotent part $N$ produces additional polynomial factors $t$ and $t^2$. 

Using the linear transformations
\begin{equation}
	x = \mu_0 q + \mu_2 \ddot{q} +\mu_4  q^{(4)}, \quad
	y = \nu_0 q + \nu_2 \ddot{q}   + \nu_4  q^{(4)}, \quad
	z= \tau_0 q + \tau_2 \ddot{q}  +\tau_4  q^{(4)},
\end{equation}	
with
\begin{align}
	\mu_0 &= -\nu_0 =  32 \lambda^3  (2 g+\lambda ), \,\, &\mu_2& =-\nu_2 =  16 \lambda (g + \lambda),  \,\,   &\mu_4& =- \nu_4 = 2 , \\
	 \tau_0 & = 32 \lambda^4, \,\, &\tau_2& = 16 \lambda^2, \,\,  &\tau_4& = 2,
\end{align}	
converts each of the equations in (\ref{thresecord}) into the threefold degenerate sixth-order PU equation (\ref{6PU})
with $\omega =  2 \lambda$. The ghostly solution then becomes
\begin{equation}
	q(t) = a_1 \sin(  \omega  t) + a_2 \cos( \omega t) + b_1  t \sin(  \omega   t) + b_2 t \cos(  \omega   t)
	+ c_1  t^2 \sin( \omega  t) + c_2 t^2 \cos( \omega  t) . \label{soldegpart}
\end{equation}
We shall see below that precisely this Jordan structure persists in the quantum theory, where it is reflected in finite chains of generalised eigenvectors organised by the hidden algebra.

\section{A hidden spectrum-generating $\mathfrak{u}(2,1)$-algebra}
We now define the intertwining operators
\begin{equation}
    a_\pm := \partial_x- \partial_y \mp \lambda (x+y), \,\, 
    b_\pm := \frac{1}{2}  \left( \partial_x + \partial_y \right)   \mp \frac{\lambda}{2} (x-y) \pm g z, \,\, 
    c_\pm :=  \partial_z    \pm g (x+ y) \mp \lambda z, 
\end{equation}
satisfying
\begin{equation}
  [H, a_\pm]= \pm 2 \lambda a_\pm, \quad
  [H, b_\pm]= \pm 2 \lambda b_\pm \mp 2 g  c_\pm  , \quad
  [H, c_\pm]=  \mp 2 g a_\pm \pm 2 \lambda  c_\pm  ,
\end{equation}
where $H$ is the quantised version of ${\cal H}$. We form the following quadratic combinations of the intertwining operators, which close under commutation
\begin{eqnarray}
	K&:=& \frac{g}{\lambda} a_+ a_- + \frac{\lambda}{g} c_+ c_- + a_+ c_- + c_+ a_- + \frac{\lambda}{g}   \left( a_+ b_- +  b_+ a_-   \right) \\
	h&:=&-\frac{1}{\lambda} \left( a_+ b_- - b_+ a_-   \right) ,  \\ 
	e&:=& \frac{1}{\lambda} \left( a_+ c_- -  c_+ a_-   \right) ,  \\ 
	f&:=& \frac{g}{\lambda^2} \left( a_+ b_- -  b_+ a_-   \right) + \frac{1}{\lambda}  \left( c_+ b_- - b_+ c_-   \right)   ,   \\ 
	v_{-2}&:=&  b_+ b_-   \\
		v_{-1}&:=& \frac{g}{\lambda} \left( a_+ b_- + b_+ a_-   \right)  +  \left( b_+ c_- + c_+ b_-   \right) ,   \\
			v_{0}&:=&  2 \frac{g^2}{ \lambda^2} a_+ a_-  + 2 c_+ c_-  - a_+ b_- - b_+ a_- + 2 \frac{g}{\lambda}   \left(  a_+ c_- + c_+ a_-    \right)   ,    \\
				v_{1}&:=&  2 \frac{g}{\lambda}  a_+ a_- + \left( a_+ c_- + c_+ a_-   \right)   ,     \\
					v_{2}&:=&  a_+ a_- . 
\end{eqnarray}
We may separate the generators into
\begin{equation}
		V_1 =\langle  e,h,f   \rangle,   \qquad 
	V_2 =\langle  v_{-2}, v_{-1},v_{0},v_{1},v_{2}  \rangle,
\end{equation}
so that, as a vector space
\begin{equation}
	\mathfrak{g} = \mathbb{R}K \oplus V_1 \oplus V_2,
\end{equation}
where $K$ is central, $V_1 \cong  \mathfrak{sl}_2  $, and $V_2$ is the irreducible 5-dimensional $\mathfrak{sl}_2 $-module. We have 
\begin{equation}
[V_1,V_1] \subset V_1,
\qquad
[V_1,V_2] \subset V_2,
\qquad
[V_2,V_2] \subset V_1.
\end{equation}
The resulting $9$-dimensional Lie algebra is isomorphic to the standard real Lie algebra $\mathfrak u(2,1)$, see for instance \cite{biedenharn1965uni}. 
Concretely we have the commutation relations
\begin{equation}
	     [h,e] = 2 e, \qquad [h,f] = -2 f, \qquad [e,f] = 2 h,    \label{sl2com}
\end{equation}
and
\begin{equation}
[h, v_m] = 2m\,v_m,  \label{hmcom}
\end{equation}
\begin{equation}
[e, v_m] = (2 - m)(3 + m)\,a_m\,v_{m+1},
\qquad
[f, v_m] = -(2 + m)(3 - m)\,a_{-m}\,v_{m-1},
\end{equation}
where $a_{-2} =1/2$, $a_{-1} =1/3$ and $a_{0} =a_{1} =-1$ due to our chosen normalisation, and
\begin{equation}
	[v_m, v_n] =2 \lambda^2  c_{m,n}^{m+n} J_{m+n}, \qquad   J_{-1}=f, \,\,\, J_{0}=h, \,\,\, J_{1}=e,
\end{equation}
for $m+n\in \{-1,0,1\}$, with all other brackets determined by antisymmetry.

The non-vanishing antisymmetric structure constants are $c_{n,m}^{n+m}=-c_{m,n}^{n+m}$, $c_{-2,1}^{-1}= c_{2,-2}^{0}= c_{-1,2}^{1}=c_{1,-1}^{0}=1$, $c_{-1,0}^{-1}=c_{0,1}^{1}=3$. The significance of this algebraic structure is that the descendant spaces $\mathcal{V}_N$
furnish finite-dimensional invariant subspaces on which $H$ acts by Jordan blocks with eigenvalue $2 \lambda N$ as we will show next.

\section{Quantum descendant spaces, $\mathfrak{sl}_2$-modules and Jordan chains}
Given the $\mathfrak{u}(2,1) $ algebraic structure we can  express the quantum Hamiltonian version $H$ of ${\cal H}$ in terms of its generators
\begin{equation}
	      H =  \frac{g}{\lambda} v_1   - \frac{g^2}{\lambda^2} v_2 -  \frac{g}{\lambda} K + 3 \lambda . 
\end{equation}
In what follows we drop the unimportant constant $3 \lambda$. We construct the vacuum state  
\begin{equation}
	\psi_0 = \exp\left[ -\frac{1}{2}   (x,y,z)G(x,y,z)^\intercal \right], \qquad G= \begin{pmatrix}
		\lambda & 0 & -g\\
		0 & -\lambda & -g\\
		-g & -g & \lambda
	\end{pmatrix},
\end{equation}
satisfying
\begin{equation}
	a_- 	\psi_0=  b_- 	\psi_0=c_- 	\psi_0= K	\psi_0= 0, \quad \Rightarrow \quad  H	\psi_0= 0. 
\end{equation}
 We notice that $\psi_0$ is not normalisable, as we expect in the resonance sector of the PU-model. We see this by noting that the matrix $G$ cannot be positive-definite for real $g$ and $\lambda$. Considering the characteristic equation for $G$
 \begin{equation}
 	 \tau ^3 -\lambda  \tau ^2 -(\lambda ^2  +2 g^2 ) \tau  +\lambda ^3=0,
 \end{equation}
 its three roots $\tau_1,\tau_2,\tau_3$ satisfy Vieta's relations
  \begin{equation}
   \tau_1 + \tau_2 + \tau_3 = \lambda, \quad
     \tau_1  \tau_2  + \tau_1  \tau_3+ \tau_2  \tau_3 = -(\lambda ^2  +2 g^2 ), \quad
       \tau_1  \tau_2  \tau_3 = \lambda^3.  \label{viete}
  \end{equation}
If all three roots were positive, then every pairwise product of the roots would be positive, so that the second equation in (\ref{viete}) is impossible to satisfy when $g, \lambda \in \mathbb{R}$.

Next we define the states
\begin{equation}
	\psi_{\ell m n}: = a_+^{\ell}c_+^{m}b_+^{n}\psi_0,
\end{equation}
and for each $N \geq 0$ the descendant spaces
\begin{equation}
\mathcal{V}_N=\operatorname{span}\left\{
	\psi_{\ell m n}:\ell + m+n =N
\right\}.
\end{equation}
Using the intertwining relations we see that the Hamiltonian $H$ maps descendant states into descendant states
\begin{equation}
H\mathcal{V}_N\subseteq \mathcal{V}_N.
\end{equation}
In particular, it acts on $\mathcal{V}_N$ as
\begin{equation}
H=2\lambda N  \mathbb{I}+M_N, 
\end{equation}
where $M_N$ is nilpotent on $\mathcal{V}_N$.  Thus every $\mathcal{V}_N$ is a generalised eigenspace of eigenvalue $2\lambda N$.
Concretely we have 
\begin{equation}
	\mathcal{V}_1=\operatorname{span}\left\{  \psi_{100}, \psi_{010} ,\psi_{001} \right\}, \label{spanV1}
\end{equation}
with
\begin{eqnarray}
	(H- 2 \lambda) \psi_{100} &=&0,  \\
		(H- 2 \lambda) \psi_{010} &=& - 2 g \psi_{100}, \quad \Rightarrow  (H- 2 \lambda)^2 \psi_{010} =0, \\
	(H- 2 \lambda) \psi_{001} &=& - 2 g \psi_{010}, \quad \Rightarrow  (H- 2 \lambda)^3 \psi_{001} =0.
\end{eqnarray}
Thus, $ \psi_{100}$ is a genuine eigenstate, whereas $ \psi_{010}$ and  $ \psi_{001}$ are rank-2 and rank-3 generalised eigenvectors. At the next level we have
\begin{equation}
	\mathcal{V}_2=\operatorname{span}\left\{  \psi_{200}, \psi_{110} , \psi_{020}, \psi_{101} , \psi_{011} , \psi_{002} \right\},
\end{equation}
with
\begin{align}
	(H- 4 \lambda) \psi_{200} &=0,  \\
	(H- 4 \lambda) \psi_{110} &= - 2 g \psi_{200}, &\Rightarrow  \,\,\,  (H- 4 \lambda)^2 \psi_{110} &= 0, \\
	(H- 4 \lambda) \psi_{020} &= - 4 g \psi_{110},  &\Rightarrow  \,\,\, (H- 4 \lambda)^3 \psi_{020} &= 0, \\
	(H- 4 \lambda) \psi_{101} &= - 2 g \psi_{110},  &\Rightarrow  \,\,\,  (H- 4 \lambda)^3 \psi_{101} &= 0, \\
	(H- 4 \lambda) \psi_{011} &= - 2 g \left( \psi_{020} +\psi_{101} \right), & \Rightarrow  \,\,\,  (H- 4 \lambda)^4 \psi_{011} &= 0, \\
		(H- 4 \lambda) \psi_{002} &= - 4 g \psi_{011}, & \Rightarrow \,\,\, (H- 4 \lambda)^5 \psi_{002} &= 0.
\end{align}
In summary, the action of $H- 4 \lambda$ on $\mathcal{V}_2$ is organised by the pattern
\begin{equation}
	\psi_{002} \to \psi_{011} \to \{\psi_{020}, \psi_{101}\} \to \psi_{110} \to \psi_{200}.
\end{equation}
The branching at level 3 suggests that there should be a better basis that directly leads to the Jordan blocks. We now show that this is indeed the case, at least partially, and we can understand the descendant spaces more systematically.

\subsection{The $\mathfrak{sl}_2$-module structure of the descendant spaces}

Denoting the distinguished $\mathfrak{sl}_2$-subalgebra inside $\mathfrak{u}(2,1)$ as
\begin{equation}
\mathfrak{s}:=\langle e,h,f\rangle \cong \mathfrak{sl}_2 ,
\end{equation}
we now describe the descendant spaces $\mathcal V_N$ as $\mathfrak{s}$-modules. As usual, by an $\mathfrak s$-module we mean a vector space $V$ on which each element of the Lie algebra $\mathfrak s$ acts linearly, in such a way that the commutator of the induced operators reproduces the Lie bracket in $\mathfrak s$. With the convention for the $\mathfrak{sl}_2$-commutation relations as in (\ref{sl2com}), and a highest-weight vector $w$ satisfying
\begin{equation}
	e w =0,\qquad h w=m w,
\end{equation}
the irreducible $\mathfrak{sl}_2$-module $D_m$ is spanned by
\begin{equation}
	w_k:=f^k w,\qquad k=0,1,\dots,m,
\end{equation}
with
\begin{equation}
	h\,w_k=(m-2k)w_k,\qquad
	f\,w_k=w_{k+1},\qquad
	e\,w_k=2k(m-k+1)w_{k-1},    \label{hfew}
\end{equation}
where $w_{-1}=w_{m+1}=0$. In particular,
\begin{equation}
	\dim D_m=m+1.
\end{equation}
Noting that the descendant space $	\mathcal V_N $ is spanned by degree-$N$ monomials in the pairwise commuting raising operators $a_+$, $b_+$,  $c_+$, it is naturally identified with a symmetric power.
It follows that, as a vector space,
\begin{equation}
	\mathcal V_N \cong \operatorname{Sym}^N(\mathcal V_1),
\end{equation}
where $\operatorname{Sym}^N$ denotes the $N$-th symmetric tensor power, see e.g. \cite{fulton2013repr} for more details.

At level $N=1$, we have $	\mathcal V_1$ as stated in (\ref{spanV1}), which is a $3$-dimensional irreducible module for $\mathfrak{s}$. Indeed, with highest-weight vector  $	w=\psi_{100}$ one finds $ew=0, \, hw=2w$, so that $\mathcal V_1 \cong D_2$.  Consequently, we have
\begin{equation}
	\mathcal V_N \cong \operatorname{Sym}^N(D_2)
\end{equation}
as $\mathfrak{s}$-modules.

Now $D_2$ is the spin-$1$ irreducible $\mathfrak{sl}_2$-module, and its symmetric powers decompose according to the standard rule
\begin{equation}
	\operatorname{Sym}^N(D_2)\cong \bigoplus_{k=0}^{\lfloor N/2\rfloor} D_{2N-4k}.
	\label{symD2}
\end{equation}

Since $\dim D_m=m+1$, this yields immediately 
\begin{equation}
	\dim \mathcal V_N
	=
	\sum_{k=0}^{\lfloor N/2 \rfloor}(2N-4k+1)
	=
	\frac{(N+1)(N+2)}{2} = \begin{pmatrix}
		N+2 \\
		2
	\end{pmatrix} ,
\end{equation}
in agreement with the number of triples $(\ell,m,n)$ satisfying $\ell+m+n=N$.

For the first few levels this gives
\begin{equation}
	\mathcal V_1\cong D_2,\qquad
	\mathcal V_2\cong D_4\oplus D_0,\qquad
	\mathcal V_3\cong D_6\oplus D_2, \qquad
		\mathcal V_4\cong D_8 \oplus D_4  \oplus D_0.
\end{equation}

We stress that this is an $\mathfrak{sl}_2$-module decomposition. Since the Hamiltonian $H$ is not an $\mathfrak{sl}_2$-intertwiner, the decomposition in (\ref{symD2}) does in general not coincide with the Jordan decomposition of $H$. Rather, it provides a basis adapted to the hidden $\mathfrak{sl}_2$-structure and the nilpotent part may still mix different irreducible $\mathfrak{sl}_2$-summands. 

Nonetheless, in some cases the $\mathfrak{sl}_2$-adapted basis also realises Jordan chains of $H$, but this need not hold for every irreducible $\mathfrak{sl}_2$-summand. In what follows we denote an $r$-dimensional Jordan block with eigenvalue $E$ as $J_r(E)$. Next we show how this structure unfolds at the lowest levels.

\subsubsection{$\mathcal V_1\cong D_2$}
The highest weight vector for $D_2$ is identified as $w= \psi_{100}$. Then $\mathcal V_1$ is spanned by the three vectors
\begin{equation}
	w_0 := w  ,\qquad
	w_1 :=  f w_0 = - \frac{2 g}{\lambda}  \psi_{100}  - 2  \psi_{010} ,  \qquad
	w_2 := f w_1 = - 4  \psi_{001} .  
\end{equation}
Thus $\mathcal{V}_1=\operatorname{span}\left\{   w_0, w_1, w_2 \right\}$ is simply a change of basis when compared with (\ref{spanV1}).
We verify that the action of the $\mathfrak{sl}_2$-generators on these vectors is indeed compatible with (\ref{hfew})
\begin{align}
	h w_0 & = 2 w_0, \qquad   & h w_1  &=0, \qquad  & h w_2 &= - 2 w_2 , \label{D21}  \\
	e w_0 & = 0, \qquad   & e w_1 & =4 w_0, \qquad  & e w_2 &= 4 w_1 ,  \\
	f w_0 & = w_1, \qquad   & f w_1 &=w_2, \qquad  & f w_2 &= 0.   \label{D23}
\end{align}	
Moreover, we find a Jordan block $J_3(2 \lambda)$
\begin{equation}
(H - 2 \lambda)^{i} w_i  \neq 0, \qquad 	(H - 2 \lambda)^{i+1} w_i =0, \qquad i=0,1,2.		
\end{equation}
Thus $\{ w_0, w_1, w_2  \}$ forms a Jordan chain of length 3 for the eigenvalue $2 \lambda$.

\subsubsection{$\mathcal V_2\cong D_4\oplus D_0$}
In this case we identify two highest weight states 
\begin{equation}
	w= \psi_{200}, \qquad \text{and} \qquad \tilde{w}=  \frac{g}{ 2 \lambda }\psi_{200}  +  \psi_{110}
	+ \frac{\lambda}{g}  \psi_{101}  + \frac{\lambda}{2 g}   \psi_{020}   ,
\end{equation}
satisfying 
\begin{equation}
	e w = 0, \,\, h w = 4 w,    \qquad \text{and} \qquad   e \tilde{w} = 0, \,\, h \tilde{w} = 0.
\end{equation}
To span $D_4$ we define the five vectors 
\begin{eqnarray}
	w_0 &:=& w  ,\\
	w_1 &:=& f w_0 = -4 \left(  \frac{g}{\lambda}  \psi_{200}  + \psi_{110}  \right), \\
	w_2&:=& f w_1 = 8 \left(   \frac{g^2}{\lambda^2}   \psi_{200}  +  \frac{2 g}{\lambda}   \psi_{110}  -\psi_{101} + \psi_{020}     \right)  \\
	w_3 &:=& f w_2 = 48 \left(   \frac{g}{\lambda}   \psi_{101}  +  \psi_{011}   \right)  \\
	w_4 &:=& f w_3 = 96 \psi_{002} .
\end{eqnarray}
The action of the $\mathfrak{sl}_2$-generators on these vectors is 
\begin{align}
	h w_0 & = 4 w_0, \quad   & h w_1  &=2 w_1 , \quad  & h w_2 &= 0 ,  \quad  & h w_3  &=- 2 w_3 , \quad  & 	h w_4 & = - 4 w_4   \\
	e w_0 & = 0, \quad   & e w_1 & =8 w_0, \quad  & e w_2 &= 12 w_1 ,  \quad  & e w_3 &= 12 w_2,  \quad    & e w_4 &= 8 w_3,  \\
	f w_0 & = w_1, \quad   & f w_1 &=w_2, \quad  & f w_2 &= w_3,   \quad  & f w_3 &= w_4,  \quad    & f w_4 &= 0. 
\end{align}	
We identify a length 5 Jordan chain with Jordan block $J_5(4 \lambda)$ entirely in $D_4$
\begin{equation}
(H - 4 \lambda)^{i} w_i  \neq 0, \qquad	(H - 4 \lambda)^{i+1} w_i =0, \qquad i=0,1,2,3,4. 		
\end{equation}
In contrast $D_0$ is spanned only by $\tilde{w}$. In this case we find 
\begin{equation}
 (H - 4 \lambda)^2 \tilde{w}  \neq 0,       \qquad	(H - 4 \lambda)^3 \tilde{w} =0 . \label{40tilde}
\end{equation}
Thus, the highest weight vector $\tilde{w}$ is not itself an eigenvector of H, but its image $\check{w}:=(H - 4 \lambda)^2 \tilde{w}  $ is an eigenvector. Noting, however, that $\check{w}= 8 g \lambda \psi_{200}$, we see that the nilpotent part $H - 4 \lambda$ maps  the $D_0$ highest weight vector into the $D_4$ summand. Instead we see that the second linearly independent state 
\begin{equation}
  \hat{w} := \psi_{020} - 2 \psi_{101}  = \frac{g^2}{\lambda^2} w_0 + \frac{g}{2 \lambda } w_1 + \frac{1}{6} w_2 - \frac{2 g}{3 \lambda }  \tilde{w}
\end{equation}
spanning the $J_1(4\lambda)$ block is not contained in either irreducible $\mathfrak{sl}_2$-summand $D_4$ or $D_0$. Rather, it is a mixed vector with nontrivial projections onto both summands. This reflects the fact that the Hamiltonian $H$ is not an $\mathfrak{sl}_2$-intertwiner, so its Jordan decomposition does not align with the decomposition into irreducible $\mathfrak{sl}_2$-modules.

\subsubsection{$\mathcal V_3 \cong D_6\oplus D_2$}
Given the decomposition, we find the two highest weight states 
\begin{equation}
	w= \psi_{300}, \qquad \text{and} \qquad \tilde{w}=  \frac{g^2}{ \lambda^2 }\psi_{300}  +  2 \psi_{201}
	+ \frac{2 g}{\lambda}  \psi_{210}  +  \psi_{120}   ,
\end{equation}
satisfying 
\begin{equation}
	e w = 0, \,\, h w = 6 w,    \qquad \text{and} \qquad   e \tilde{w} = 0, \,\, h \tilde{w} = 2 \tilde{w}.
\end{equation}
The seven vectors that span $D_6$ are
\begin{eqnarray}
	w_0 &:=& w  ,\\
	w_1 &:=& f w_0 = -6  \left(  \frac{g}{\lambda}  \psi_{300}  + \psi_{201}  \right), \\
	w_2&:=& f w_1 = 12 \left(   \frac{2 g^2}{\lambda^2}   \psi_{300}  - \psi_{201}  + \frac{ 4 g}{\lambda}     \psi_{210} + 2 \psi_{120}     \right) , \\
	w_3 &:=& f w_2 =  48 \left(  \frac{ g^3}{\lambda^3}   \psi_{300}  - \frac{3 g }{\lambda} \psi_{201}  + \frac{ 3 g^2}{\lambda^2}     \psi_{210} + \frac{3 g}{\lambda}\psi_{120}  - 3  \psi_{111} + \psi_{030}    \right) , \quad \\
	w_4 &:=& f w_3 =  -288  \left(   \frac{2 g^2}{\lambda^2}   \psi_{201}  - \psi_{102}  + 2    \psi_{021} + \frac{4 g }{\lambda}\psi_{111}     \right),  \\
	w_5 &:=& f w_4 = -2880  \left(  \frac{g}{\lambda}  \psi_{102}  + \psi_{012}  \right), \\
	w_6 &:=& f w_5 = - 5760\,\psi_{003}. 
\end{eqnarray}
We recover the action in (\ref{hfew}) on these states with $m=6$. Moreover, we find a Jordan block $J_7(6 \lambda)$
\begin{equation}
(H - 6 \lambda)^{i} w_i \neq 0, \qquad	(H - 6 \lambda)^{i+1} w_i =0, \qquad i=0,1,\ldots,6. 		
\end{equation}
The $D_2$ space is spanned by the three vectors
\begin{eqnarray}
	\tilde{w}_0 &:=& \tilde{w}  ,\qquad  \\
	\tilde{w}_1 &:=&  f \tilde{w}_0 =  -2 \left(  \frac{g^3}{\lambda^3}   \psi_{300} +\frac{2g}{\lambda}    \psi_{201}  + \frac{3 g^2}{ \lambda^2}    \psi_{210}  
	+ \frac{3 g}{ \lambda}    \psi_{120}  + 2  \psi_{111} +  \psi_{030}     \right),\\
	\tilde{w}_2 &:=& f \tilde{w}_1 =-4\left(  \frac{g^2}{\lambda^2}   \psi_{201} + 2 \psi_{102} + \psi_{021} + \frac{2 g}{ \lambda}    \psi_{111}  
	\right) .  
\end{eqnarray}
The action of the $\mathfrak{sl}_2$-generators on these vectors is the same as in (\ref{D21})-(\ref{D23}).

In this case we find
\begin{equation}
  (H - 6 \lambda)^{i+2} \tilde{w}_i  \neq 0 , \qquad	(H - 6 \lambda)^{i+3} \tilde{w}_i =0, \qquad i=0,1,2. 		\label{60tilde}
\end{equation}
This suggests introducing $\check w_i:=(H-6\lambda)^{i+2}\tilde w_i $. Formally these satisfy the annihilation conditions expected for a length-3 Jordan chain
\begin{equation}
(H-6\lambda)^{i}\check w_i \neq 0,      \quad(H-6\lambda)^{i+1}\check w_i=0,\qquad i=0,1,2.
\end{equation}
However, we find that $\check w_i \propto  \psi_{300}$ for $i=0,1,2$, so that they are not linearly independent and moreover they do not lie in $D_2$. Instead we find that the missing $J_3(6 \lambda)$ block is built from vectors belonging to both summands $D_6$ and $D_2$
\begin{eqnarray}
	\hat{w_0} &=&  \psi_{120} - 2 \psi_{201} = \frac{g^2}{\lambda^2} w_0 + \frac{g}{3 \lambda} w_1 + \frac{1}{15} w_2 + \frac{3}{5} \tilde{w}_0 ,  \notag	\\
	  	\hat{w_1} &=& \frac{1}{g} \left[   \psi_{111} - \frac{1}{2} \psi_{030} \right]=
	  	\frac{g^2}{2 \lambda^3} w_0 + \frac{g}{4 \lambda^2} w_1 + \frac{1}{15 \lambda} w_2  + \frac{1}{120 g} w_3 
	  	- \frac{1}{10 \lambda} \tilde{w}_0 + \frac{1}{20 g} \tilde{w}_1 , \qquad \,\,	\\
	  		\hat{w_2} &=&   \frac{1}{2 g^2}  \left[  \frac{1}{2} \psi_{021} - \psi_{102} \right]=  
	  		-\frac{1}{10 \lambda g}   \left[  \frac{g}{24 \lambda} w_2 +  \frac{1}{48} w_3  + \frac{\lambda}{ 144 g  } w_4 - \frac{g}{\lambda}      \tilde{w}_0 - \frac{1}{2}   \tilde{w}_1 -   \frac{3 \lambda }{8 g}   \tilde{w}_2               \right], \notag
\end{eqnarray}
satisfying
\begin{equation}
	(H - 6 \lambda)^{i} \hat{w}_i \neq 0, \qquad	(H - 6 \lambda)^{i+1} \hat{w}_i =0, \qquad i=0,1,2. 		
\end{equation}

Although the descendant spaces admit a clean decomposition into irreducible  $\mathfrak{sl}_2$-modules, this decomposition does not in general coincide with the Jordan decomposition of the Hamiltonian. The nilpotent part of H mixes different  $\mathfrak{sl}_2$ summands, so the hidden algebra organises the representation content of the resonance sector, but does not by itself determine the Jordan normal form.

\section{From Lie symmetries to a tri-Hamiltonian formulation}	

Next we identify Lie point symmetries of the dynamical flow and use them to
construct a tri-Hamiltonian formulation of the system.

Using the previously introduced phase-space vector $\vec{q}$ the equations of motion (\ref{thresecord}) can also be written as 
\begin{equation}
	\dot{ \vec{q} } = \vec{V}(\vec{q}), \label{Hflow}
\end{equation}
where the differential operator
\begin{eqnarray}
	V &=&	\dot x \partial_x  + \dot y  \partial_y +  \dot z \partial_z 
	+ \left[8 g\lambda z-4\lambda^2 x -4 g^2 (x+y)  \right]  \partial_{\dot{x}}  +
	\left[ 4 g^2 (x+y)-8 g\lambda z-4\lambda^2 y \right] \partial_{\dot{y}} \notag \\ 
	&&  +[ 8g \lambda(x+y)- 4\lambda^2 z]  \partial_{\dot{z}}  
\end{eqnarray}
acts like a vector field when interpreted as a derivative on $C^\infty (\mathbb{R}^6) $.
Next we consider infinitesimal transformations generated by 
\begin{equation}
	X = \sum_{i=1}^6 \xi_i(\vec{q})  \partial_{q_i}  ,
\end{equation}
which are symmetries of the flow if and only if
\begin{equation}
	[X,V]=0.
 \end{equation}
Here $[X,V]$ denotes the Lie bracket of the vector fields $X$ and $V$ on
$\mathbb{R}^6$. Making a linear Ansatz for the components $ \xi_i (\vec{q})  $, we find six linearly independent solutions for the Lie bracket to vanish
\begin{eqnarray}
	X_1 &=& \frac{1}{2} \left[  \dot{z}   (  \partial_x-  \partial_y  )  + ( \dot{x}+\dot{y})\partial_z +4  \lambda  (z \lambda -2 g
	(x+y)) ( \partial_{\dot{y}}- \partial_{\dot{x} } )  -4  (x+y) \lambda ^2 \partial_{\dot{z}} \right],   \quad \\
	X_2 &=&  \frac{1}{2} \left[  z( \partial_x  - \partial_y) +  (x+y)  \partial_z + \dot{z}(  \partial_{\dot{x}} -\partial_{\dot{y}} )  +  (\dot{x}+ \dot{y})   \partial_{\dot{z}}  \right], \\
	X_3 &=&     \frac{1}{2}  (  \dot{x}  +  \dot{y}  )   (  \partial_y - \partial_x  )     +2 \lambda ^2 (x+y)  (  \partial_{\dot{x} }  - \partial_{\dot{y} }   ) ,\\
	X_4 &=&  \frac{1}{2} \left[ (2 x+y) \partial_x    -  x   \partial_y  + z  \partial_z + (2  \dot{x} +   \dot{y} )  \partial_{\dot{x} }   -  \dot{x}   \partial_{\dot{y} } +   \dot{z} \partial_{\dot{z} }   \right] ,  \\
	X_5 &=&   \frac{1}{2} \left[ (x+y)   (  \partial_y - \partial_x  ) + (\dot{x}+ \dot{y})     (  \partial_{\dot{y} }  - \partial_{\dot{x} }   )   \right] ,     \\
	X_6 &=&  \frac{1}{2} \left[  (2 \dot{x}  + \dot{y} )    \partial_x  -  \dot{x}     \partial_y   +  \dot{z}     \partial_z \right]    + 2 \left[ (2 g z \lambda -g^2 (x+y)-(2 x+y) \lambda ^2)\partial_{\dot{x}}      \right.  \\
	&&   \qquad \qquad \qquad  \left. +  (g^2 (x+y)-2 g z \lambda +x \lambda ^2)     \partial_{\dot{y}}     +    \lambda  (2 g (x+y)-z \lambda ) \partial_{\dot{z}}                  \right]   .  \notag
\end{eqnarray}
All of the $X_i$ are mutually commuting, i.e. $ [ X_i,X_j] =0 $ for $i,j=1, \dots,6$.  Note that $2 X_3 + 2 X_6 = V$.

Alternatively the flow in (\ref{Hflow}) can also be produced from 
\begin{equation}
	\frac{d\vec{x}}{dt} = J \nabla {\cal H } ,  \label{Hamflow}
\end{equation}
with $J$ denoting the standard canonical Poisson tensor, $\vec{x} =(x,y,z, p_x,p_y,p_z)^\intercal $ and keeping in mind that $p_x= \dot{x}/2$, $p_y= -\dot{y}/2$, $p_z= \dot{z}/2$. Since the $X_i$ are symmetries of the flow, it is natural to ask whether they generate alternative Hamiltonian descriptions of the same dynamics. Acting with suitable linear combinations of these symmetry generators on the original Hamiltonian ${\cal H }_1:={\cal H }$, we obtain two further Hamiltonians ${\cal H }_2$ and ${\cal H }_3$
\begin{eqnarray}
	{\cal H }_2  &:=& X_4({\cal H }_1)+	X_5({\cal H }_1)-X_2({\cal H }_1) \\
	&=& p_x^2-p_y^2-2 p_x p_z+2 p_y p_z+p_z^2+(g+\lambda )^2 x^2+\left(g^2+2 g \lambda -\lambda ^2\right) y^2+\lambda ^2 z^2 \notag \\
	&&+2 g (g+2 \lambda ) x y-2 \lambda  (2
	g+\lambda ) x z-2 \lambda  (2 g+\lambda ) y z ,   \notag \\
	{\cal H }_3  &:=& X_4({\cal H }_2)+	X_5({\cal H }_2)+X_2({\cal H }_2) \\
	&=&2 p_x p_y-2 p_y^2+p_z^2+g^2 x^2+\left(g^2-2 \lambda ^2\right) y^2+ \lambda ^2 z^2+\left(2 g^2-2 \lambda ^2\right) x y-4 g \lambda  x z-4 g \lambda  y z ,\notag 
\end{eqnarray}
Indeed with the Poisson tensors
\begin{equation}
	J_1 = \begin{pmatrix} 0 & 0 & 0 & 1 & 0 & 0 \\
		0 & 0 & 0 & 0 & 1 & 0 \\
		0 & 0 & 0 & 0 & 0 & 1 \\
		-1 & 0 & 0 & 0 & 0 & 0 \\
		0 & -1 & 0 & 0 & 0 & 0 \\
		0 & 0 & -1 & 0 & 0 & 0    \end{pmatrix}, \,\,
	J_2 = \begin{pmatrix} 0 & 0 & 0 & 2 & 1 & 1 \\
		0 & 0 & 0 & -1 & 0 & -1 \\
		0 & 0 & 0 & 1 & 1 & 1 \\
		-2 & 1 & -1 & 0 & 0 & 0 \\
		-1 & 0 & -1 & 0 & 0 & 0 \\
		-1 & 1 & -1 & 0 & 0 & 0   \end{pmatrix}, \,\,
	J_3 = \begin{pmatrix} 0 & 0 & 0 & 2 & 1 & 0 \\
		0 & 0 & 0 & -1 & 0 & 0 \\
		0 & 0 & 0 & 0 & 0 & 1 \\
		-2 & 1 & 0 & 0 & 0 & 0 \\
		-1 & 0 & 0 & 0 & 0 & 0 \\
		0 & 0 & -1 & 0 & 0 & 0  \end{pmatrix},\quad
\end{equation}
we generate the same flow as in (\ref{Hamflow}). Since both $J_2$ and $J_3$ are constant antisymmetric matrices, they automatically satisfy the Jacobi identity and hence define Poisson structures. Thus we have obtained a tri-Hamiltonian system. It is easily verified that these Hamiltonians are mutually in involution, i.e. $ \{{\cal H }_1,{\cal H }_2\}=\{ {\cal H }_1,{\cal H }_3\}=\{{\cal H }_2,{\cal H }_3\}=0$. In general, we find that $X_1$, $X_3$ and $X_6$ are symmetries of all three Hamiltonians 
\begin{equation}
	X_1({\cal H }_i)= X_3({\cal H }_i)=X_6({\cal H }_i)=0, \qquad i=1,2,3 . 
\end{equation}
The other symmetry operators act as 
\begin{align}
	X_2({\cal H }_1) &= {\cal H }_1 -{\cal H }_2,   \quad  &  X_4({\cal H }_1) &= 2 {\cal H }_1 -{\cal H }_3, \quad    &X_5({\cal H }_1) &= - {\cal H }_1 +{\cal H }_3,  \\
	X_2({\cal H }_2) &= {\cal H }_3 -{\cal H }_2,      &X_4({\cal H }_2) &= {\cal H }_1 +{\cal H }_2 -{\cal H }_3,    &X_5({\cal H }_2) &= - {\cal H }_1 +{\cal H }_3,  \\    
	X_2({\cal H }_3) &= {\cal H }_1 -{\cal H }_2,     & X_4({\cal H }_3) & = {\cal H }_1 ,    &X_5({\cal H }_3) &= - {\cal H }_1 +{\cal H }_3  .
\end{align}
Thus the symmetry generators $X_2,X_4,X_5$ close on the three-dimensional space spanned by ${\cal H }_1,{\cal H }_2,{\cal H }_3$, while $X_1,X_3,X_6$ annihilate all three Hamiltonians.

Remarkably, the quantum versions of the alternative Hamiltonians ${\cal H }_2$ and ${\cal H }_3$ also lie in the span of the hidden algebra generators $K,v_1,v_2$, 
\begin{eqnarray}
	H_2 &=&  \left( 1+\frac{g}{\lambda} \right) v_1   - \left(  \frac{g^2}{\lambda^2} +2 \frac{g}{\lambda}  \right)    v_2 -  \frac{g}{\lambda} K + 3 \lambda  ,         \\ 
	H_3 &=&  \frac{g}{\lambda}  v_1   + \left(1-   \frac{g^2}{\lambda^2}  \right)    v_2 -  \frac{g}{\lambda} K + 3 \lambda ,
\end{eqnarray}
so that the quantum tri-Hamiltonian structure is naturally encoded by the same $\mathfrak u(2,1)$ framework. The additive constant is dynamically irrelevant and will be suppressed from now on. Thus, in the remainder of this section, $H$ denotes the shifted operator $H-3\lambda$.

Note that the generators $X_i$ belong to the Lie algebra of infinitesimal symmetries of the classical phase-space flow on $\mathbb R^6$, whereas the operators $a_\pm,b_\pm,c_\pm$ and the derived generators $K,e,h,f,v_m$ act on wavefunctions on configuration space. Hence these two sets of generators need not admit a direct identification, even though they arise from the same underlying dynamical system.

\section{An apparent enhanced quantum symmetry}

Next we address the question of whether there exist additional charges that commute with all quantum Hamiltonians $H_1,H_2,H_3$. The common conservation of such charge can be understood without referring to
the explicit differential operator realisation. Since the three Hamiltonians differ only
by elements of the abelian subspace $\langle v_1,v_2\rangle$, up to the
central element $K$, their common quantum symmetries within $U(\mathfrak u(2,1))$ are therefore
naturally sought in the centraliser of $\langle v_1,v_2\rangle$. 

This centraliser problem is graded by the adjoint action of $h$. We have 
$[h,v_1]=2v_1$, $[h,v_2]=4v_2$, thus, if $Q=\sum_r Q^{(r)}$ is decomposed into $h$-weight components,
$[h,Q^{(r)}]=r Q^{(r)}$, then $[Q^{(r)},v_1]$ and $[Q^{(r)},v_2]$ have weights $r+2$ and $r+4$, respectively. Hence the centraliser equations $[Q,v_1]=[Q,v_2]=0$ can be solved independently in each $h$-weight sector.

At quadratic order, the first nontrivial centraliser element beyond the
obvious Hamiltonian generators occurs in the weight-$+2$ sector. We
therefore make the most general quadratic ansatz in this sector, modulo
trivial additions from $\langle v_1,v_2,K\rangle$, and impose the centraliser equations. This yields
\begin{equation}
	Q = e h + 4 e - \frac{1}{3 \lambda^2} v_0 v_1 - \frac{2}{\lambda^2} v_{-1} v_2 .
\end{equation}
By construction this operator commutes with $v_1$, $v_2$, and $K$, and hence with $H_1,H_2,H_3$. However, it is not an independent new symmetry. Indeed, we find
\begin{eqnarray}
	Q& =& -\frac{2 (2 g+\lambda ) \left(g^2+g \lambda +\lambda ^2\right)}{3 \lambda ^5} { H}_1^2 
	-\frac{2 g}{3 \lambda ^3}   { H}_2^2     -\frac{4 g^3}{3 \lambda ^5}      { H}_3^2  + \frac{2 \left(3 g^2+2 g \lambda +\lambda ^2\right)}{3 \lambda ^4} { H}_1  { H}_2   \qquad  \label{QinH} \\
	&& + \frac{2 g \left(4 g^2+3 g \lambda +2 \lambda ^2\right)}{3 \lambda ^5}    { H}_1  { H}_3 
	-\frac{2 g^2}{\lambda ^4}  { H}_2  { H}_3 .  \notag 
\end{eqnarray}
Thus $Q$ belongs to the commutative algebra generated by the three Hamiltonians. It is therefore a common quantum symmetry only in the dependent sense and does not provide an additional independent conserved charge.

There is a small subtlety in comparing this statement with the classical limit. One has to distinguish the commutative
classical symbol of a quantum symmetry from the full ordered symbol of its
differential operator representative. Expanding the ordered operator $Q$ before passing to phase
space produces lower-order terms, including terms linear in the momenta. These
terms arise from derivatives acting on coordinate-dependent coefficients and
therefore depend on the chosen ordering. They are quantum ordering corrections.
The classical object relevant for functional independence and superintegrability
is instead obtained by first passing the hidden-algebra generators to their
commutative classical symbols and then forming the corresponding polynomial. In this manner we find the classical counterpart to $Q$ as
\begin{equation}
	Q_{cl} = e_{cl}  h_{cl}  - \frac{1}{3 \lambda^2} (v_0)_{cl}  (v_1)_{cl}  - \frac{2}{\lambda^2} (v_{-1})_{cl}  (v_2)_{cl} ,
\end{equation}
where the classical symbols are obtained by replacing $\partial_w \rightarrow i p_w, \, w \rightarrow w$ for $w=x,y,z$. Note that the term $4e$ is a purely ordering contribution. Using $[h,e]=2e$, we may write $ 4e=2[h,e]=2he-2eh$. 
In the commutative classical limit the distinction between $he$ and $eh$
disappears, so this commutator contribution vanishes $2h_{\rm cl}e_{\rm cl}-2e_{\rm cl}h_{\rm cl}=0$.
Thus $4e$ does not contribute to the classical commutative expression for the charge. 

For the representation in (\ref{QinH}) we simply replace the quantum Hamiltonians $H_i$ by their classical counterparts ${\cal H}_i$. 

We note that the quadratic Casimir 
\begin{equation}
C_2=  h^2+ \{e,f\} + \frac{1}{3 \lambda^2}  v_0^2 +\frac{1}{\lambda^2}  \{ v_{-1}, v_1    \} +\frac{2}{\lambda^2}  \{ v_{-2} ,v_2    \}
\end{equation}
of the hidden $\mathfrak u(2,1)$ algebra evidently also commutes with $H_1,H_2,H_3$, since the Hamiltonians lie in the span of
$K,v_1,v_2$ and $C_2$ commutes with all generators of the algebra. This
conservation is, however, central and therefore automatic. It should be
distinguished from the charge $Q$, which is not central in $U(\mathfrak u(2,1))$, but is nevertheless reducible because it is a polynomial in the Hamiltonians. but belongs specifically to the centraliser of the Hamiltonian subspace.

\section{Filtration structure of $H_2$, $H_3$ and $Q$}

Next we observe that formally the action of $H_2$ and $H_3$ on the $\mathfrak{sl}_2$-modules have the same filtration-compatible nilpotency pattern as \(H_1\) on the \(\mathfrak{sl}_2\)-adapted chains.
$H=H_1$. For instance, in the cases in which the summand of the $\mathfrak{sl}_2$-decomposition is identical to the Jordan block $J_r(\kappa \lambda)$ we find
\begin{equation}
	\left( H_j - \kappa \lambda    \right)^i w_i \neq 0, \qquad 	\left( H_j - \kappa \lambda    \right)^{i+1}  w_i = 0,
	\qquad i=0,1, \ldots ,r-1, \,\,\, j=1,2,3.
\end{equation}  
Similarly the formal nilpotency pattern of $H_2$ and $H_3$ on the $\mathfrak{sl}_2$-adapted chains is the same as for $H_1$.

This pattern can be understood from the fact that
\begin{equation}
H_1,H_2,H_3\in -\frac{g}{\lambda}K+\operatorname{span}\{v_1,v_2\},
\end{equation}
i.e. the three Hamiltonians differ only by linear combinations of the operators $v_1$ and $v_2$, while $K$ contributes only a scalar shift. Since by (\ref{hmcom}) $v_1$ and $v_2$ belong to the spin-2 multiplet $V_2$ with
\begin{equation}
[h,v_1]=2v_1,\qquad [h,v_2]=4v_2,
\end{equation}
their action on a highest-weight basis $w_i=f^i w_0$ is triangular
\begin{equation}
v_1 w_i \in \operatorname{span}\{w_0,\dots,w_{i-1}\},\qquad
v_2 w_i \in \operatorname{span}\{w_0,\dots,w_{i-2}\}.
\end{equation}
Hence all three Hamiltonians preserve the same filtration associated with the $\mathfrak{sl}_2$-chain, which explains why their formal nilpotent action on these chains is identical. The difference between $H_1,H_2$ and $H_3$ only becomes visible once different $\mathfrak{sl}_2$-summands mix, and this is precisely where their Jordan decompositions differ.
In the $\mathcal V_2$ case the $J_1(4\lambda)$ blocks are built on different linearly independent states
\begin{eqnarray}
	 \hat{w} &:=&  \frac{g}{g+\lambda} \psi_{110} +\psi_{101} - \frac{1}{2} \psi_{020} =
	 -\frac{g^2 (g+ 3 \lambda)}{2 \lambda^2 (g+ \lambda)} w_0 - \frac{g (g+ 2 \lambda)}{4 \lambda (g+ \lambda)} w_1 -\frac{1}{12} w_2 + \frac{g}{ 3 \lambda} \tilde{w},	 \qquad \\
	 \hat{w} &:=&   -\frac{\lambda}{g} \psi_{110} +\psi_{101} - \frac{1}{2} \psi_{020} = \left(
	1-  \frac{ g^2}{2 \lambda^2 } \right)w_0 + \left( \frac{\lambda}{ 4g} - \frac{g}{4 \lambda}  \right)w_1 -\frac{1}{12} w_2 + \frac{g}{ 3 \lambda} \tilde{w},	
\end{eqnarray}
for $H_2$ and $H_3$, respectively. Thus, just as $H_1$,  the other two tri-Hamiltonian partners also mix different  $\mathfrak{sl}_2$ summands. We find a similar behaviour for $\mathcal V_3$.

Finally we report how the enhanced symmetry $Q$ acts on the lowest descendant spaces. By construction
the state $Qw_i$ must lie in the weight space with weight $m-2i+2$. If this weight space is one-dimensional, $Qw_i$ is necessarily proportional to $w_{i-1}$. If several irreducible $\mathfrak{sl}_2$-summands contain the same weight, $Q$ may mix them.

For  $\mathcal V_1\cong D_2 $, using the basis  $w_0,w_1,w_2 $ introduced above, we find
\begin{equation}
	Qw_0=0,\qquad
	Qw_1=\frac{40}{3}w_0,\qquad
	Qw_2=-\frac{40}{3}w_1 .
\end{equation}
Hence, on  $\mathcal V_1 $, the enhanced charge acts as a weight raising operator along the single  $\mathfrak{sl}_2 $-chain. Since there is only one state at each relevant weight, no mixing between different  $\mathfrak{sl}_2 $-summands can occur.

The first nontrivial behaviour appears in  $\mathcal V_2\cong D_4\oplus D_0 $. Letting  $w_0,\ldots,w_4 $ denote the  $D_4 $-chain and let  $\tilde w $ span the  $D_0 $-summand, we compute
\begin{equation}
	Qw_0 =0, \quad Qw_1 =\frac{112}{3}w_0, \,\, Qw_2 =\frac{56}{3}w_1, \,\,  
	Qw_3 =-\frac{56}{3}w_2-\frac{1792g}{3\lambda}\tilde w, \,\,  Qw_4 =-\frac{112}{3}w_3,
\end{equation}
and
\begin{equation}
	Q\tilde w=\frac{14\lambda}{3g}w_1 .    \label{qwtilde}
\end{equation}
Thus  $Q $ preserves the full descendant space  $\mathcal V_2 $, but it does not preserve the irreducible summands  $D_4 $ and  $D_0 $ separately. The term proportional to  $\tilde w $ in  $Qw_3 $, together with the fact that  $Q\tilde w $ lies in  $D_4 $, shows explicitly that  $Q $ mixes the two  $\mathfrak{sl}_2 $-modules.

This behaviour is consistent with the general interpretation of  $Q $. It is a common symmetry of the quantum Hamiltonians and therefore maps generalised energy eigenspaces into themselves. However, since it has nonzero  $\mathfrak{sl}_2 $-weight, it is not an  $\mathfrak{sl}_2 $-intertwiner. Rather, it acts as an additional weight raising-type operator inside the finite descendant spaces, preserving  $\mathcal V_N $ while generally mixing the irreducible  $\mathfrak{sl}_2 $-summands appearing in its decomposition.

Moreover, we have 
\begin{equation}
	Q^{i+1}w_i=0,
	\qquad i=0,\ldots,2 \quad \text{on } \mathcal V_1,
	\qquad
	i=0,\ldots,4 \quad \text{on } D_4\subset\mathcal V_2 .
\end{equation}
This nilpotency is a direct consequence of the fact that repeated applications of  $Q $ move states upward in the finite  $\mathfrak{sl}_2 $-weight filtration until no higher-weight state is available.

For the additional  $D_0 $-highest-weight vector  $\tilde w\subset\mathcal V_2 $, the situation illustrates the difference between an irreducible  $\mathfrak{sl}_2 $-summand and the full descendant-space filtration. Although $\tilde w$ is the highest-weight vector of the $D_0$-summand, it is not annihilated by $Q$. Instead, as seen in (\ref{qwtilde}), $Q\tilde w$ lies in the $D_4$-summand. Nevertheless  $Q $ remains nilpotent on the full space, and in this case $Q^3\tilde w=0$.

\section{No-go theorem for positive-definite Hamiltonians}

In \cite{felski2026three}  it was shown that, for the corresponding non-resonant model, one can construct a positive-definite linear combination of the tri-Hamiltonians that generates the same flow. We now show that this is impossible for the present resonant system. Defining the linear combinations
\begin{equation}
	\bar{J} = c_1 J_1 + c_2 J_2 + c_3 J_3, \qquad   	\bar{{\cal H}} = c_4 {\cal H}_1 + c_5 {\cal H}_2 + c_6 {\cal H}_3,
\end{equation}
we compute 
\begin{eqnarray}
	\bar{J}  \nabla	\bar{{\cal H}} &=&   2 \left[ c_2 \left(c_4+c_6\right)- \left(c_1+c_3\right) c_5\right]X_1 (\vec{x}) +   2 \left(c_1+c_2+c_3\right) \left(c_4+c_5+c_6\right)   X_6 (\vec{x}) \qquad  \\&& +  2\left[   c_3 c_6+c_2 \left(c_5+c_6\right)+c_1 \left(c_4+c_5+2 c_6\right)   \right] X_3 (\vec{x})   . \notag
\end{eqnarray}
Recalling that $2 X_3 + 2 X_6 = V$, we obtain the same flow as in (\ref{Hflow}) by setting the coefficients in front of $X_1$ to zero and those multiplying $X_3 $ and $X_6$ to 2. Solving these equations yields
\begin{equation}
	c_4 = \frac{c_1^2+c_3 c_1-c_2 c_3}{\rho ^3}, \qquad
	c_5 =\frac{c_2}{\rho^2}, \qquad 
	c_6 = \frac{\bar{\rho} }{\rho ^3},
\end{equation}
where $\rho=c_1 + c_2 + c_3 \neq 0$, $\bar{\rho} =c_1 \left(c_2+c_3\right)+c_3 \left(2 c_2+c_3\right)    $ and $c_1,c_2,c_3$ remain free parameters. For these choices we can write
\begin{equation}
	\bar{{\cal H}} = \vec{p}^\intercal   M_p  \vec{p}  + \vec{x}^\intercal   M_v  \vec{x} , \quad \text{with}\,\,
	\vec{p} = (p_x,p_y,p_z)^\intercal , \,\,  \vec{x} = (x,y,z)^\intercal 
\end{equation}
with
\begin{eqnarray}
	M_p &=&   \begin{pmatrix}
		\frac{\rho ^2-\bar{\rho }}{\rho ^3} & \frac{\bar{\rho }}{\rho ^3} & -\frac{c_2}{\rho ^2} \\
		\frac{\bar{\rho }}{\rho ^3} & \frac{c_1 \rho +c_2^2-2 \rho ^2}{\rho ^3} & \frac{c_2}{\rho ^2} \\
		-\frac{c_2}{\rho ^2} & \frac{c_2}{\rho ^2} & \frac{1}{\rho } 
	\end{pmatrix}  ,  \\ 
	M_v &=&    \begin{pmatrix}
		\frac{\rho  \left(2 c_2 g \lambda +\rho  \left(g^2+\lambda ^2\right)\right)-\lambda ^2 \bar{\rho }}{\rho ^3} & \frac{g \rho  \left(2 c_2 \lambda +g \rho
			\right)-\lambda ^2 \bar{\rho }}{\rho ^3} & -\frac{\lambda  \left(c_2 \lambda +2 g \rho \right)}{\rho ^2} \\
		\frac{g \rho  \left(2 c_2 \lambda +g \rho \right)-\lambda ^2 \bar{\rho }}{\rho ^3} & \frac{-\left(\lambda ^2 \left(\bar{\rho }+\rho ^2\right)\right)+2 c_2 g
			\lambda  \rho +g^2 \rho ^2}{\rho ^3} & -\frac{\lambda  \left(c_2 \lambda +2 g \rho \right)}{\rho ^2} \\
		-\frac{\lambda  \left(c_2 \lambda +2 g \rho \right)}{\rho ^2} & -\frac{\lambda  \left(c_2 \lambda +2 g \rho \right)}{\rho ^2} & \frac{\lambda ^2}{\rho }
	\end{pmatrix}   ,
\end{eqnarray}
up to an irrelevant additive constant. A direct computation yields
\begin{equation}
	\Delta_2(M_p) = -\frac{1}{\rho^2} , \qquad \Delta_2(M_v) = -\frac{\lambda^4}{\rho^2}, \qquad 
	\det M_p  = - \frac{1}{\rho^3} ,   \qquad  \det M_v = - \frac{ \lambda^6 }{ \rho^3} , \label{detmin}
\end{equation}
where $\Delta_2(M) $ denotes the second leading principal minor of the matrix $M$. Recalling Sylvester's criterion, stating that a real symmetric matrix is positive-definite if and only if all its leading principal minors are positive, we deduce from (\ref{detmin}) that neither $M_p$ nor $M_v$ can be positive-definite. Since positivity of $\bar {\cal H}$ would require both quadratic forms to be positive-definite, no positive-definite linear combination of the tri-Hamiltonians can generate the same flow.

\section{Conclusions}

We have investigated a ghostly three-dimensional Hamiltonian system that realises
the fully degenerate resonant sixth-order PU oscillator. At the
classical level the resonance manifests itself through a non-diagonalisable
phase-space flow: the evolution matrix decomposes into two complex-conjugate
Jordan blocks of length three. This directly explains the appearance of secular
terms proportional to $t $ and $t^2$ in the solutions of the corresponding
sixth-order PU equation.

At the quantum level we have shown that the resonant model possesses a hidden
spectrum-generating algebra. Starting from three pairs of first-order
intertwining operators, we constructed quadratic combinations that close into a
nine-dimensional Lie algebra isomorphic to $\mathfrak u(2,1)$. This algebra
contains a distinguished $\mathfrak{sl}_2$-subalgebra and a five-dimensional
irreducible $\mathfrak{sl}_2$-module. It therefore provides the natural
extension of the hidden $\mathfrak{su}(2)$ structure found previously in the
fourth-order resonant PU model \cite{fring2026spectrum}.

The hidden algebra organises the finite-dimensional descendant spaces generated
from the formal vacuum. Each space $\mathcal V_N$ is invariant under the
Hamiltonian and carries a nontrivial Jordan structure with eigenvalue
$2\lambda N$. We showed that these spaces admit a clean decomposition into
irreducible $\mathfrak{sl}_2$-modules (\ref{decompVN}). However, this representation-theoretic decomposition does not in general
coincide with the Jordan decomposition of the Hamiltonian. The nilpotent part
of the Hamiltonian can mix different irreducible $\mathfrak{sl}_2$-summands,
so that the hidden algebra organises the resonance sector but does not by
itself diagonalise the Jordan structure.

We also constructed a tri-Hamiltonian formulation of the same classical flow.
Using linear Lie point symmetries, we identified three Hamiltonians and three
constant Poisson tensors that generate identical equations of motion. Their
quantum counterparts are naturally encoded in the same hidden
$\mathfrak u(2,1)$ algebra, lying in the span of $K,v_1,v_2$, up to
irrelevant additive constants. Thus, although multi-Hamiltonian formulations already occur away from
resonance, at the fully resonant point the tri-Hamiltonian partners are
naturally encoded in the same hidden $\mathfrak u(2,1)$ algebra that
organises the quantum Jordan structure.

In addition, we analysed a higher-order element $Q$ in the common centraliser of the three Hamiltonians. Although this operator commutes with $H_1,H_2,H_3$, it is not an independent quantum symmetry, as it can be expressed as a quadratic polynomial in the Hamiltonians themselves. Its commutative classical counterpart is likewise functionally dependent on the classical tri-Hamiltonians and therefore does not provide an additional integral of motion. Consequently, $Q$ neither establishes classical superintegrability nor provides a genuinely new quantum integral. Rather, it is a reducible symmetry reflecting the algebraic closure of the tri-Hamiltonian family inside $U(\mathfrak u(2,1))$. This also clarifies the role of ordering corrections: while they are present in the differential-operator representative of $Q$, they do not turn $Q$ into an algebraically independent quantum charge.

Finally, we proved that the positive-definite Hamiltonian mechanism available
in the corresponding non-resonant three-dimensional model breaks down at the
fully degenerate resonant point. For every linear combination of the three
Hamiltonians that still generates the original flow, the kinetic and potential
quadratic forms fail Sylvester's criterion for positive-definiteness. Hence no
positive-definite Hamiltonian arises within this tri-Hamiltonian family.

Taken together, these results show that the fully resonant sixth-order PU model
is not merely a limiting case of the non-resonant theory. It is a genuinely
singular system in which higher-order Jordan structure, hidden
$\mathfrak u(2,1)$ symmetry, a reducible higher-order centraliser element and
multi-Hamiltonian geometry coexist. The construction also suggests several
directions for future work. It would be natural to investigate whether analogous
hidden algebras appear in still higher-order resonant PU systems, and whether
their structure follows a systematic hierarchy. Another important question is
whether the formal non-normalisable quantum states encountered here can be
related, perhaps through a non-unitary similarity transformation \cite{fring2025ghos} or a modified
inner product, to a fully consistent Hilbert-space realisation. These questions
are directly tied to the broader problem of understanding ghost degrees of
freedom and resonance phenomena in higher time-derivative theories.

\medskip

\noindent {\bf Acknowledgments}: IM was supported by the Australian Research Council Future Fellowship FT180100099. 

\newif\ifabfull\abfulltrue

\end{document}